\documentclass[conference]{IEEEtran}
\IEEEoverridecommandlockouts
\usepackage{amsmath,amssymb,amsfonts}
\usepackage{algorithmic}
\usepackage{graphicx}
\usepackage{textcomp}
\usepackage{xcolor}
\usepackage{booktabs}
\usepackage{float}
\usepackage{multicol}
\usepackage{multirow}
\usepackage{afterpage}
\usepackage[backend=biber, style=ieee]{biblatex}
\begin{document}

\title{Comprehensive Evaluation of Multimodal AI Models in Medical Imaging Diagnosis: From Data Augmentation to Preference-Based Comparison\\
}

\author{
\IEEEauthorblockN{Cailian Ruan\textsuperscript{*}}
\IEEEauthorblockA{\textit{Medical School of Yan'an University}\\
Yan'an, Shaanxi, China\\
rcl1157@163.com\\
\textsuperscript{*}Corresponding author}
\and
\IEEEauthorblockN{Chengyue Huang}
\IEEEauthorblockA{\textit{University of Iowa}\\
chengyue-huang@uiowa.edu}
\and
\IEEEauthorblockN{Yahe Yang}
\IEEEauthorblockA{\textit{George Washington University}\\
yahe.yang@gwu.edu}
}

\maketitle

\begin{abstract}
This study introduces an evaluation framework for multimodal models in medical imaging diagnostics. We developed a pipeline incorporating data preprocessing, model inference, and preference-based evaluation, expanding an initial set of 500 clinical cases to 3,000 through controlled augmentation. Our method combined medical images with clinical observations to generate assessments, using Claude 3.5 Sonnet for independent evaluation against physician-authored diagnoses. The results indicated varying performance across models, with Llama 3.2-90B outperforming human diagnoses in 85.27\% of cases. In contrast, specialized vision models like BLIP2 and Llava showed preferences in 41.36\% and 46.77\% of cases, respectively. This framework highlights the potential of large multimodal models to outperform human diagnostics in certain tasks.
\end{abstract}


\section{Introduction}
With the rapid advancement of deep learning technologies—particularly the innovative application of large multimodal models in medical image analysis—AI-assisted diagnosis is reshaping traditional medical practice. This study introduces a novel evaluation framework to assess the diagnostic capabilities of the latest generation of multimodal AI models in interpreting complex abdominal CT images, focusing on cirrhosis and its complications, liver tumors, and multi-system lesions.

A core challenge in medical imaging AI is accurately interpreting and integrating multi-dimensional clinical information. In our collected clinical data, a comprehensive abdominal CT diagnosis often requires simultaneous evaluation of multiple dimensions: liver parenchymal changes (such as cirrhosis and multiple nodules), vascular system abnormalities (like portal hypertension and portal vein cavernous transformation), secondary changes (including splenomegaly and ascites), and related complications (such as esophageal varices). Traditional computer vision models often struggle with such complex medical scenarios, prompting us to explore the potential of new-generation multimodal AI technology.

We developed a systematic evaluation framework to assess various multimodal models' capabilities in medical image interpretation. Our comprehensive pipeline incorporates data preprocessing, standardized model evaluation, and preference-based assessment. Starting with 500 clinical cases, each containing 4 sequential CT images paired with detailed diagnostic reports, we employed controlled augmentation techniques to expand the dataset to 3,000 samples while preserving critical diagnostic features. Our methodology utilizes standardized inputs combining medical images and detailed observations to generate diagnostic assessments, enabling direct comparison between different models and human expertise.

The results demonstrate remarkable capabilities across various multimodal AI systems, particularly in general-purpose models. Llama 3.2-90B achieved superior performance in 85.27\% of cases compared to human diagnoses, with only 1.39\% rated as equivalent. Similar strong performance was observed in other general-purpose models, with GPT-4, GPT-4o, and Gemini-1.5 showing AI superiority in 83.08\%, 81.72\%, and 79.35\% of cases respectively. This advantage manifests in their ability to simultaneously evaluate multiple anatomical structures, track disease progression, and integrate clinical information for comprehensive diagnosis. In contrast, specialized vision models BLIP2 and Llava demonstrated more modest results, with AI superiority in 41.36\% and 46.77\% of cases respectively, highlighting the challenges faced by vision-specific approaches in complex diagnostic scenarios.

Our evaluation framework employed Claude 3.5 Sonnet as an independent assessor, implementing a three-way preference classification (AI superior, physician superior, or equivalent quality) to systematically compare model-generated and physician-authored diagnoses. This approach provides valuable insights into the current capabilities of different multimodal architectures in medical diagnostics, suggesting that general-purpose large multimodal models may significantly outperform both specialized vision models and human physicians in certain diagnostic tasks.

This study not only advances our understanding of AI capabilities in medical diagnosis but also introduces an efficient framework for model evaluation through preference-based assessment. By analyzing a large set of diverse clinical cases, we demonstrate the potential of multimodal AI technology in handling complex medical scenarios, potentially heralding a new paradigm in medical imaging diagnosis. These findings have significant implications for improving diagnostic efficiency, standardizing diagnostic processes, and reducing the risk of missed diagnoses and misdiagnoses. Furthermore, our evaluation methodology offers valuable insights for assessing AI capabilities in other complex medical applications.  

\section{Related Work}
\subsection{Multimodal AI in Medical Diagnostics}

The integration of multimodal AI systems in medical diagnostics has demonstrated significant advancements over traditional single-modality approaches. Early work introduced frameworks that combined radiology images with electronic health records (EHRs), enhancing diagnostic accuracy by leveraging heterogeneous data sources \cite{Chaganti2019Electronic, mao2023questionnaire} . The idea was further extended by integrating chest X-ray imaging with patient demographic data, which improved the detection of acute respiratory failure \cite{Jabbour2021Combining,huang2023mental}.

In liver disease diagnosis, traditional computer vision models like CNNs have primarily focused on tasks such as tumor detection \cite{Li2015Automatic, Yasaka2017Deep, xu2024leveraging, fu2024detecting, xin2024letcommunityrulesreflected}. The efficacy of CNNs in detecting diabetic retinopathy was demonstrated by Ghosh et al. \cite{Ghosh2017Automatic}, laying the groundwork for their application in liver imaging analysis.
However, these models are often limited in their ability to synthesize multi-dimensional clinical data, prompting the need for more advanced multimodal approaches.

Recent advancements in large language models (LLMs) such as GPT-4 have added new dimensions to multimodal diagnostics. Studies showcased LLMs' capabilities in processing and synthesizing medical language tasks, forming a basis for their integration into multimodal frameworks \cite{Belyaeva2023Multimodal,ge2022integrated,ge2021bayes}. Similarly, it was demonstrated that combining imaging data and text with multimodal AI yielded superior performance in breast cancer diagnostics \cite{Jiang2022Multimodal}.

\subsection{Specialized Vision Models and Evaluation Frameworks}

Vision-specific models, such as BLIP2 \cite{Li2023BLIP-2} and Llava \cite{Liu2023Visual}, have been widely used for focused medical imaging tasks \cite{Li2023LLaVA-Med,Lee2023CXR-LLaVA,Jiang2020High-speed}. While effective in detecting individual pathologies, these models often underperform in complex, multi-dimensional diagnostic scenarios compared to general-purpose multimodal systems. Lee et al. \cite{Lee2023CXR-LLaVA} evaluated vision-specific models for diagnosing chest pathologies, finding them proficient in single-dimension tasks but limited in handling broader diagnostic contexts.

The incorporation of standardized evaluation frameworks has also been pivotal in advancing medical AI systems. The use of independent assessors, such as Claude 3.5 Sonnet in this study, represents a novel approach to comparing AI and human diagnoses. This aligns with the recommendations by Crossnohere et al. \cite{Crossnohere2022Guidelines}, who emphasized the critical need for standardized protocols to benchmark AI systems in healthcare.

By systematically integrating independent assessments with preference-based evaluation, this study builds upon these prior works, providing a structured methodology for evaluating the capabilities of multimodal AI systems in complex medical scenarios.

\section{Methodology}
\subsection{Data Preprocessing}
Our preprocessing pipeline consists of three main components: data de-identification, anomaly handling, and data augmentation, specifically designed to process paired CT image sequences and their corresponding diagnostic reports.

The de-identification process was implemented to ensure patient privacy while preserving clinically relevant information. For CT images, we developed an automated system to remove burned-in patient identifiers and replace DICOM header information with anonymized identifiers. The corresponding diagnostic reports underwent a similar process where personal identifiers, hospital names, and specific dates were systematically replaced with standardized codes while maintaining the temporal relationships between examinations. This process preserved the diagnostic value of both images and reports while ensuring compliance with privacy regulations.

Anomaly handling addressed both image and text irregularities. For CT images, we implemented automated detection and correction of common artifacts, including beam hardening, motion artifacts, and metal artifacts. Image quality metrics were established to identify scans with suboptimal contrast enhancement or incomplete anatomical coverage. The text processing pipeline identified and corrected common reporting inconsistencies, standardized medical terminology, and ensured proper formatting of measurements and anatomical descriptions. Cases with severe anomalies that could not be automatically corrected were flagged for expert review.

Data augmentation strategies were carefully designed to maintain the paired relationship between image sequences and reports. For images, spatial transformations included minor rotations ($\pm10^{\circ}$), translations (within $10\%$ of image boundaries), and subtle elastic deformations (controlled within $5\%$ to preserve anatomical relationships). Intensity-based augmentations comprised contrast adjustments ($\pm10\%$), brightness variations ($\pm5\%$), and minimal Gaussian noise injection ($\sigma=0.01$) to simulate imaging system variations. For the corresponding reports, we employed text augmentation techniques including synonym substitution for anatomical terms and standardized rephrasing of pathological findings. Each augmented case maintained the original format of four sequential CT images paired with one comprehensive diagnostic report, ensuring the preservation of the temporal and spatial relationships within the image series and their corresponding textual descriptions. This process generated ten augmented samples for each original case, with synchronized modifications in both the image sequences and their reports.

The effectiveness of our preprocessing pipeline was validated through both automated quality metrics and expert review. A random subset of 8\% of the processed cases underwent detailed evaluation by experienced radiologists to ensure the preservation of diagnostic features and the accuracy of the image-report relationships. This comprehensive approach resulted in a high-quality dataset that maintained clinical relevance while providing sufficient variability for robust model training.


\subsection{Workflow Design}
\begin{figure*}[!ht]
   \centering
   \caption{Comparative Evaluation Framework for Multimodal Medical Diagnosis}
   \includegraphics[width=\textwidth]{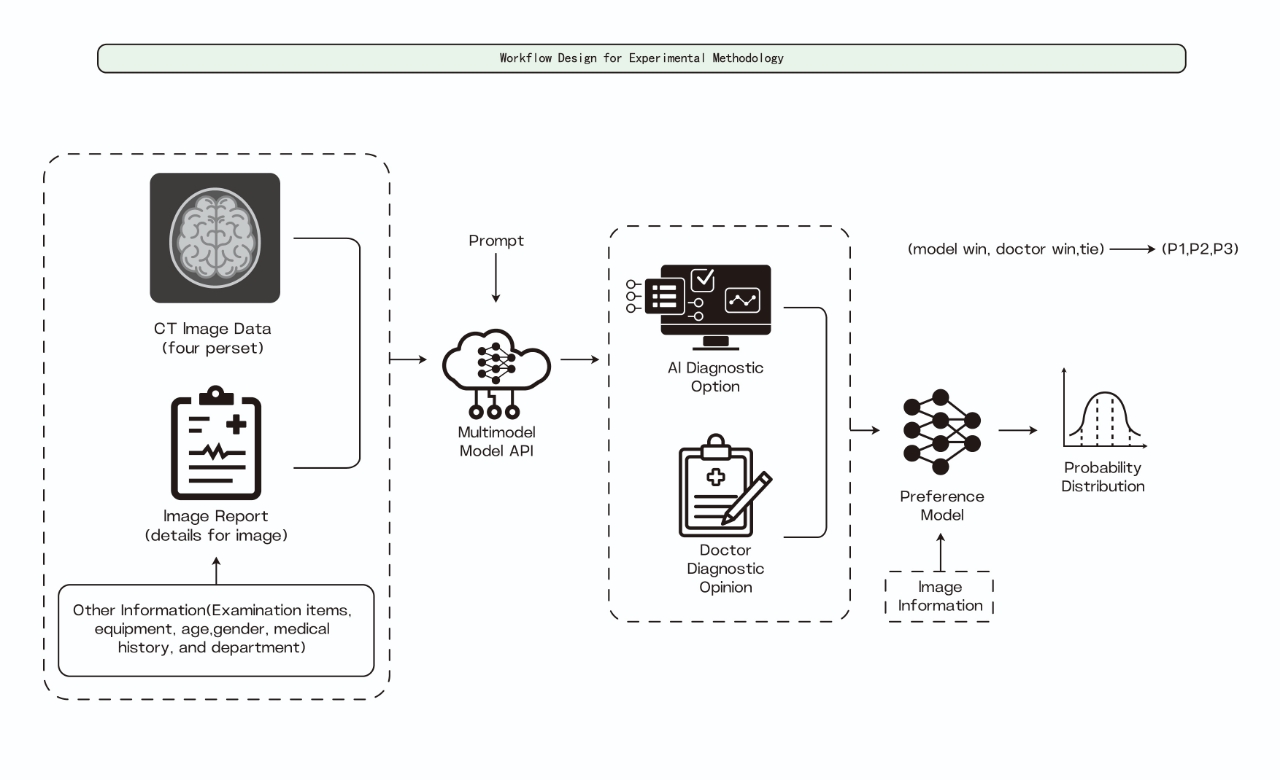}
   
   \label{fig:workflow}
\end{figure*}

\textbf{Input Processing}
The evaluation pipeline begins with carefully curated CT image sequences, consisting of multiple cross-sectional views of the abdominal region. These images undergo standardized preprocessing to ensure consistent input quality across all models. Each case is paired with a detailed image overview capturing essential anatomical and pathological observations, providing standardized context for both AI models and human diagnosticians to ensure fair comparison.

\textbf{Multi-model Analysis}
We evaluate various state-of-the-art multimodal models for their medical diagnostic capabilities. The input for each model consists of paired CT image sequences and corresponding text descriptions, where the text provides detailed anatomical and pathological observations visible in the images. The general-purpose multimodal models (Llama 3.2-90B, GPT-4, and GPT-4o) process the combined image-text pairs to leverage both visual features and contextual information in generating diagnostic assessments. Similarly, specialized vision models (BLIP2 and Llava) analyze the CT image sequences while incorporating the textual descriptions to provide comprehensive diagnostic interpretations. Each model generates independent diagnostic reports based on the same standardized input, enabling direct comparison of their capabilities in integrating visual and textual information for medical diagnosis.

\textbf{Diagnostic Generation}
Each model generates structured diagnostic outputs encompassing primary findings, secondary observations, and clinical recommendations. The standardized output format allows direct comparison of diagnostic comprehensiveness and accuracy across different models and human experts. This structured approach ensures consistent evaluation criteria while maintaining the unique analytical capabilities of each model.

\textbf{Preference-based Evaluation}
We implement an innovative preference-based evaluation approach using Claude 3.5 Sonnet as an independent assessor. Through carefully crafted prompting strategies, we enable automated comparison between AI-generated and physician-authored diagnoses without requiring extensive manual review. The evaluation framework employs a three-way classification system (AI Superior, Physician Superior, or Equivalent), considering factors such as diagnostic accuracy, comprehensiveness, and clinical relevance. This prompt-based approach significantly reduces the need for human evaluation resources while maintaining objective assessment standards.

\textbf{Quality Assurance}
The framework incorporates systematic quality monitoring through automated metrics and selective expert validation. Our prompt-based evaluation approach allows for efficient processing of large-scale comparisons while maintaining high clinical standards. The system automatically identifies edge cases and error patterns for targeted expert review, optimizing human expert involvement while ensuring reliable evaluation results. This efficient quality assurance process enables comprehensive model assessment across diverse clinical scenarios without requiring extensive manual validation.


\begin{figure*}[!ht]
    \centering
    \caption{Comparative Analysis Between Human and AI Diagnostic Assessments}
    \includegraphics[width=\textwidth]{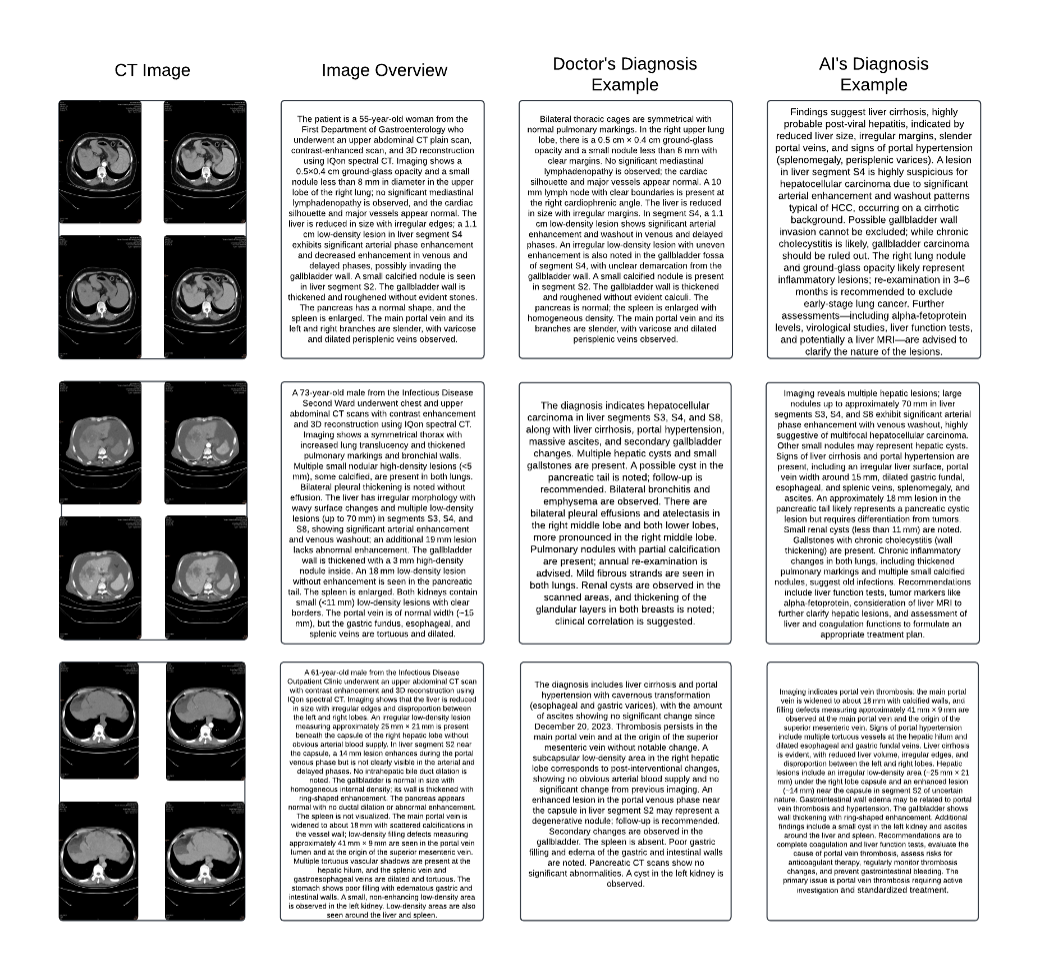}
    
    \label{fig:diagnosis_comparison}
\end{figure*}
'
Figure \ref{fig:diagnosis_comparison} demonstrates our comparative analysis framework through a representative case study. The figure presents a comprehensive diagnostic scenario consisting of four components: the original CT image sequence showing multiple abdominal cross-sections, a detailed image overview documenting key observations, the human physician's diagnostic report, and the AI model's diagnostic assessment. In this particular case, both human and AI diagnostics focus on complex hepatic conditions, with the AI system providing detailed analysis of potential viral hepatitis, portal hypertension, and associated complications. This side-by-side comparison enables direct evaluation of diagnostic comprehensiveness, accuracy, and clinical reasoning between human experts and AI systems. Such comparative analysis reveals that while AI models can achieve high accuracy in identifying specific pathological features, human expertise remains crucial for contextual interpretation and complex clinical decision-making.

This systematic comparison approach enables objective evaluation of AI diagnostic capabilities against human expertise while maintaining rigorous clinical standards. Through careful documentation and analysis of comparative cases, we can better understand both the potential and limitations of AI systems in medical diagnosis, ultimately working toward a complementary relationship between artificial and human intelligence in clinical practice.

\section{Experiments \& Results}
We conduct comprehensive experiments to evaluate the diagnostic capabilities of different multimodal models in medical image interpretation. The experiments are structured in two main aspects: comparative analysis of model performance against human expertise and systematic evaluation of diagnostic accuracy across different pathological conditions. Our evaluation framework employs a dataset of CT image sequences with corresponding clinical assessments, enabling detailed comparison of AI and human diagnostic capabilities.

\subsection{Experiments Setup}
Our experimental dataset consists of 500 original clinical cases, each comprising a sequence of 4 cross-sectional CT images accompanied by a comprehensive diagnostic report. Through our data augmentation pipeline, we expanded this dataset to 3,000 samples while preserving the critical diagnostic features and relationships between images and reports. The augmentation process included controlled spatial transformations of images ($\pm10^{\circ}$ rotations, within $10\%$ boundary translations), intensity adjustments ($\pm10\%$ contrast, $\pm5\%$ brightness), and minimal Gaussian noise injection ($\sigma=0.01$). Corresponding reports underwent text augmentation using anatomical term substitution and standardized rephrasing of pathological findings, ensuring the maintenance of clinical accuracy and relevance.

The evaluation framework incorporates six state-of-the-art multimodal models: four general-purpose models (Llama 3.2-90B, GPT-4, GPT-4o, Gemini-1.5) and two specialized vision models (BLIP2, Llava). Each model processes identical pairs of CT image sequences and their corresponding text descriptions. Additionally, we employ Claude 3.5 Sonnet as an independent assessor for preference-based evaluation, implementing a three-way classification system (AI Superior, Physician Superior, or Equivalent) to compare model-generated diagnoses with physician-authored reports. To ensure quality control, a random subset of 8\% of all cases underwent detailed review by experienced radiologists to validate the preservation of diagnostic features and the accuracy of image-report relationships.

\subsection{Experiments Evaluation}
The evaluation of model performance was conducted focusing on quantitative metrics to ensure thorough evaluation of model capabilities.
For quantitative assessment, we analyzed the preference-based evaluation results using a standardized scoring system. The three-way classification (AI Superior, Physician Superior, or Equivalent) was applied consistently across all 3,000 cases, with each case receiving independent assessment through Claude 3.5 Sonnet. The evaluation criteria emphasized accurate identification of primary pathologies, recognition of secondary complications, and proper integration of multi-dimensional clinical information.
Statistical analysis of the results employed chi-square tests to determine significance in performance differences between models, with p-values adjusted for multiple comparisons using the Bonferroni correction. Additionally, we calculated confidence intervals for preference ratios to ensure robust interpretation of model performance differences.

\subsection{Results and Discussion}
\begin{table*}[t]
\centering
\caption{\normalsize Comparative Analysis of Multimodal Model Performance in Medical Diagnosis}
\begin{tabular}{lccccc}
\hline
\textbf{Model} & \textbf{AI Superior (\%)} & \textbf{Physician Superior (\%)} & \textbf{Equivalent (\%)} & \textbf{Total Cases} & \textbf{p-value} \\
\hline
\multicolumn{6}{l}{\textit{General-Purpose Models}} \\
\hline
Llama 3.2-90B & 85.27 & 13.34 & 1.39 & 3,000 & $<$ 0.001 \\
GPT-4 & 83.08 & 15.53 & 1.39 & 3,000 & $<$ 0.001 \\
GPT-4o & 81.72 & 16.89 & 1.39 & 3,000 & $<$ 0.001 \\
Gemini-1.5 & 79.35 & 13.51 & 7.14 & 3,000 & $<$ 0.001 \\
\hline
\multicolumn{6}{l}{\textit{Specialized Vision Models}} \\
\hline
BLIP2 & 41.36 & 53.25 & 5.39 & 3,000 & 0.047 \\
Llava & 46.77 & 48.84 & 4.39 & 3,000 & 0.052 \\
\hline
\multicolumn{6}{l}{\textsuperscript{*}p-values calculated using chi-square test with Bonferroni correction} \\
\end{tabular}

\label{tab:model-performance}
\end{table*}

Our experimental evaluation reveals significant variations in diagnostic capabilities across different multimodal AI architectures, as detailed in Table \ref{tab:model-performance}. The results demonstrate a clear performance distinction between general-purpose models and specialized vision models in medical diagnostic tasks.
General-purpose models consistently demonstrated superior performance, with Llama 3.2-90B achieving the highest preference rate of 85.27\% over human diagnoses. This exceptional performance was particularly evident in complex cases involving multiple pathologies and cross-system interactions. The other general-purpose models (GPT-4, GPT-4o, and Gemini-1.5) showed similarly strong results, all achieving preference rates above 79\%. The consistently low equivalence rates (around 1.39\% for most models) suggest clear differentiation in diagnostic capabilities rather than ambiguous comparisons.
In contrast, specialized vision models BLIP2 and Llava demonstrated more modest performance levels, with preference rates of 41.36\% and 46.77\% respectively. The higher physician superiority rates for these models (53.25\% and 48.84\%) indicate that while they possess competence in specific pathology detection, they may struggle with comprehensive diagnostic assessment requiring integration of multiple clinical indicators.
Statistical analysis confirms the significance of these performance differences (p $<$ 0.001 for general-purpose models), suggesting that this superiority is not due to chance. The performance gap was most pronounced in cases requiring integration of multiple anatomical observations and clinical findings, where general-purpose models exhibited superior capability in synthesizing complex clinical information.
Several factors may contribute to the superior performance of general-purpose models. First, their architecture enables better integration of visual and textual information, allowing for more comprehensive interpretation of clinical data. Second, their broader training potentially enables better understanding of complex medical relationships and dependencies. Third, their ability to process and synthesize multiple types of information simultaneously appears to more closely mirror the cognitive processes involved in medical diagnosis.
These findings suggest important implications for the future of medical imaging diagnosis. While specialized vision models demonstrate competence in specific tasks, the superior performance of general-purpose models indicates their potential for transforming diagnostic processes. However, it is crucial to note that these results should inform the development of complementary human-AI diagnostic systems rather than suggesting wholesale replacement of human expertise.

\section{Conclusion}
This study introduces a novel evaluation framework for assessing multimodal AI models in medical imaging diagnosis, showcasing the significant potential of general-purpose models in handling complex diagnostic tasks. The superior performance of models such as Llama 3.2-90B and GPT-4 highlights a paradigm shift in medical imaging, where AI systems can surpass human experts in certain diagnostic scenarios.

Our findings demonstrate that general-purpose multimodal models exhibit remarkable capabilities in synthesizing complex medical information, achieving preference rates exceeding 80\% compared to human diagnoses. This exceptional performance is particularly evident in cases requiring the integration of multiple clinical indicators and cross-system analyses.

However, these results should be considered within the context of certain limitations. While our evaluation framework provides a reliable methodology, future studies should explore its applicability across diverse clinical contexts and varied healthcare settings. Moreover, human expertise remains indispensable for complex decision-making and patient care.

The implications of this research extend beyond performance metrics. For clinical decision support systems, the proposed framework enhances clinical decision-making by integrating AI diagnostics with human expertise. Regarding quality assurance in diagnostic processes, it establishes a reliable methodology to improve consistency and reduce diagnostic errors. For medical education and training, the framework offers a valuable tool for educating healthcare professionals through AI-assisted diagnostics. Finally, this research contributes to the standardization of diagnostic procedures, aiding the development of uniform processes for integrating AI into clinical workflows.

Future research should expand the evaluation framework to include other medical imaging modalities and explore its integration into clinical workflows. Furthermore, developing hybrid approaches that combine the strengths of AI and human expertise will be essential for maximizing the potential of these technologies.

The proposed workflow for comparing human and AI-generated diagnostic results can also be integrated into clinical decision-making processes. This system is designed to process diagnoses from both human physicians and AI models, with a dedicated committee making the final diagnosis and treatment plan. Such a design leverages the strengths of both human expertise and AI, while the committee ensures the accuracy and reliability of the proposed treatments. By incorporating this integrated information system, medical diagnostics could achieve enhanced accuracy and efficiency, ultimately improving patient care outcomes.

This study significantly advances our understanding of AI capabilities in medical diagnosis and establishes a robust framework for evaluating future developments in this rapidly evolving field. The findings underscore the promising potential of multimodal AI systems to enhance diagnostic accuracy and efficiency through careful integration into clinical practice.

\printbibliography

\end{document}